\documentclass[aps,prb,preprint,groupedaddress,superscriptaddress,showpacs]{revtex4-1}
\usepackage{longtable}
\usepackage{graphicx}

\begin{document}

\title{Phonon and magnetic dimer excitations in Fe-based S=2 spin ladder compound BaFe$_2$Se$_2$O} %

\author{Z. V. Popovi\'c}
\affiliation{Center for Solid State Physics and New Materials, Institute of Physics Belgrade, University of Belgrade, Pregrevica 118, 11080 Belgrade, Serbia}
\author{M. \v S\'cepanovi\'c}
\affiliation{Center for Solid State Physics and New Materials, Institute of Physics Belgrade, University of Belgrade, Pregrevica 118, 11080 Belgrade, Serbia}
\author{N. Lazarevi\'c}
\affiliation{Center for Solid State Physics and New Materials, Institute of Physics Belgrade, University of Belgrade, Pregrevica 118, 11080 Belgrade, Serbia}
\author{M. M. Radonji\'c}
\affiliation{Scientific Computing Laboratory, Institute of Physics Belgrade, University of Belgrade, Pregrevica 118, 11080 Belgrade, Serbia}
\author{D. Tanaskovi\'c}
\affiliation{Scientific Computing Laboratory, Institute of Physics Belgrade, University of Belgrade, Pregrevica 118, 11080 Belgrade, Serbia}
\author{Hechang Lei$^{\S}$}
\affiliation{Condensed Matter Physics and Materials Science Department, Brookhaven National Laboratory, Upton, New York 11973-5000, USA}
\author{C. Petrovic}
\affiliation{Condensed Matter Physics and Materials Science Department, Brookhaven National Laboratory, Upton, New York 11973-5000, USA}

\begin{abstract}
Raman scattering spectra of new Fe-based S=2 spin ladder compound BaFe$_2$Se$_2$O are measured in a temperature range between 15 K and 623 K. All six A$_{1g}$ and two B$_{1g}$ Raman active modes of BaFe$_2$Se$_2$O,  predicted by the factor-group analysis, have been experimentally observed at energies that are in a rather good agreement with the lattice dynamics calculation. The antiferromagnetic long-range spin ordering in BaFe$_2$Se$_2$O below $T_N$=240 K leaves a fingerprint both in the A$_{1g}$ and B$_{1g}$ phonon mode linewidth and energy. In the energy range between 400 and 650 cm$^{-1}$ we have observed magnetic excitation related structure in the form of magnon continuum, with the peaks corresponding to the singularities in the one dimensional density of magnon states. The onset value of magnetic continuum (2$\Delta_{S}$) is found at about 437 cm$^{-1}$ at 15 K. The magnetic continuum disappears at about 623 K, which lead us to conclude that the short-range magnetic ordering in BaFe$_2$Se$_2$O exists apparently up to 2.6$T_N$.
\end{abstract}

\date{\today}
\maketitle

\section{Introduction}
Iron based superconductors are among the top priorities in physics, since their discovery in 2008. \cite{Kamihara} Within this area one of the most active fields in research are ternary alkali metal iron selenide superconductors. The progress in this field is recently summarized in Ref. \onlinecite{Dagotto}. While searching for new Fe-chalcogenide (S, Se, Te) compounds that are related to high temperature superconductors in their structural and electronic properties, another intriguing class of materials has been discovered\cite{Dagotto} - two-leg ladder selenides: BaFe$_2$Se$_3$ and BaFe$_2$Se$_2$O. In fact, iron-based compounds can have not only superconducting but also low-dimensional magnetic properties\cite{Dagotto, Lei, Han} (forming spin chains, spin-ladders, spin-dimers, etc.), as in the case of cuprates \cite{ZoranSr14} or vanadates. \cite{Milan} This opens up a new field of research in the direction of synthesis of new Fe-based materials, and the new physics of the S = 2 quantum spin systems\cite{Kabbour}.

BaFe$_2$Se$_3$ is an insulator with a long-range antiferromagnetic (AFM) order with $T_N$ around 250 K and short-range AFM order at higher temperatures\cite{LeiSe3, Saparov, Caron}. It was shown that a dominant order involves 2$\times$2 blocks of ferromagnetically aligned iron spins, whereas these blocks order antiferromagnetically, in the same manner as the block-AFM $\surd5\times\surd5$ state of the iron-vacancy ordered A$_2$Fe$_4$Se$_5$.\cite{Ye,NenadKFeSe,Nenadmagnoni}

\begin{figure}
\includegraphics[width=0.5\textwidth]{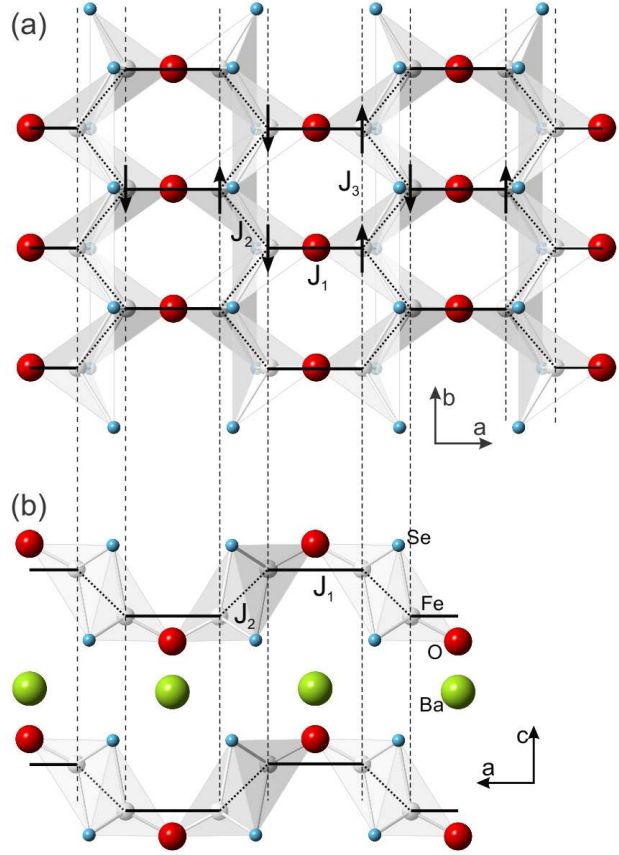}
\caption{(Color online) Schematic representation of BaFe$_2$Se$_2$O crystal structure. (a) Fe-Se(O) layer in the (ab) plane. (b) the (ac) plane of BaFe$_2$Se$_2$O crystal structure. $J_1$ represents the AFM Fe-O-Fe exchange interaction, $J_2$ is the AFM inter-dimer Fe-Se-Fe exchange interaction. $J_3$ is ferromagnetic Fe-Se-Fe exchange interaction along the ladder legs. Arrows denote orientation of the magnetic moments (spins). }
\label{fig1}
\end{figure}

BaFe$_2$Se$_2$O belongs to the family of the layered iron oxychalcogenides, which have been recently synthesized \cite{Lei, Han, Zhu, Cekic, Zhao}. Actually, this is the first layered iron-oxychalcogenide with an alkali-earth element. The band structure calculations of BaFe$_2$Se$_2$O revealed\cite{Han} the narrowing of Fe 3\emph{d} bands near the Fermi energy, which leads to the localization of magnetism and the Mott insulating behavior. Besides the long-range AFM transition at 240 K, which is confirmed with the resistivity, specific heat and magnetic susceptibility measurements \cite{Lei, Han}, there are two additional transitions at about 115 K and 42 K, where magnetic susceptibility drops abruptly \cite{Lei}. The magnetic entropy up to 300 K is much smaller than the expected value for Fe$^{2+}$ in tetrahedral crystal fields, and M\"{o}ssbauer spectrum indicates that long range magnetic order is unlikely at 294 K. Both results suggest that a short range magnetic correlations exist above the room temperature \cite{Lei}.

Barium iron oxyselenide has unique crystal structure which is built up by stacking the Ba cations and Fe-Se(O) layers alternatively along the \emph{c}-axis [Fig.1(b)]. Within the Fe-Se(O) layers, double chains of edge-shared Fe-Se(O) tetrahedra propagate along the Se atoms parallel to the \emph{b}-axis. The Fe-Se(O) double chains are bridged by oxygen along the \emph{a}-axis [Fig.1(a)]. The crystal structure is orthorhombic, space group $Pmmn$ with two formula units (Z=2) per unit cell.

In BaFe$_2$Se$_2$O, all iron ions are in the Fe$^{2+}$ oxidation state and in the high spin S = 2 state\cite{Kabbour}. The magnetic structure of BaFe$_2$Se$_2$O may be considered as the S=2 spin ladder system, where AFM exchange interaction $J_1$ along the rungs is dominant (strong Fe-O-Fe superexchange interaction). According to the local-spin-density-approximation (LSDA) calculations\cite{Han} the inter-ladder interaction $J_2$ is more than three times weaker than along the rungs. The exchange interaction $J_3$ along the ladder legs can be neglected\cite{Han}, see Fig. 1. Because the magnetic exchange interaction is short range and the magnetic coupling further than the second nearest neighbor is almost equal to zero, this magnetic system can be described as two-dimensional AFM dimer system with non-negligible inter-dimer interaction.

So far there are no data about the phonon properties of this compound. In this paper we have measured polarized Raman scattering spectra of BaFe$_2$Se$_2$O in the temperature range between 15 K and 623 K (350 $^0$C). Besides the optical phonons, which are assigned using polarized measurements and the lattice dynamical calculations, we have observed magnon continuum, which is related to the AFM dimer structure. Spin gap value of 27 meV is obtained from the position of the onset of the magnetic continuum. High temperature Raman scattering measurements shows that the short-range magnetic order disappears at about 623 K=2.6$T_N$. In the short-range order region we have observed coupling of antiferromagnetic spin fluctuations with the B$_{1g}$ symmetry phonons.

\section{Experiment and numerical method}

Single crystals of BaFe$_2$Se$_2$O were grown using self-flux method, described in details elsewhere.\cite{Lei} Raman scattering measurements were performed on (001)-oriented samples in backscattering micro-Raman configuration. Low temperature measurements were performed using KONTI CryoVac continuous flow cryostat coupled with JY T64000 Raman system. High temperature measurements were done using LINKAM THMS600 heating stage coupled with the TriVista 557 Raman system. In order to avoid oxidation of the sample the high temperature measurements were done in Ar inert atmosphere. The 514.5 nm and 488 nm lines of an Ar$^{+}$/Kr$^{+}$ mixed gas laser were used as excitation sources.

The calculations are performed within the density functional theory as implemented in the QUANTUM ESPRESSO package\cite{QE}. The lattice dynamics is calculated within the density functional perturbation theory\cite{Baroni} using the projector augmented-wave method with the PBE exchange-correlation functional which is used to obtain ultra-soft pseudo-potentials. The Brillouin zone is sampled with a Monkhorst-–Pack 16x16x10 \textbf{k}-space mesh. Unit cell is constructed using experimental values of the lattices parameters\cite{Lei} ($Pmmn$ space group with unit cell parameters: \emph{a} = 0.98518 nm, \emph{b} = 0.41332 and \emph{c} = 0.67188 nm). Fractional coordinates are relaxed until all forces acting on individual atoms became smaller than 10$^{-4}$ Ry/a.u. and they agree with experimental coordinates within a few percent. The energy cutoffs for the wave functions and the electron densities are 64 Ry and 782 Ry, respectively, which are the highest suggested radii for the chosen pseudo-potentials.

\section{Results and discussion}

The BaFe$_2$Se$_2$O unit cell consists of two formula units comprising of 12 atoms. The site symmetries of atoms in $Pmmn$ space group are C$_{2v}$ (Ba, O) and C$_{s}$ (Fe, Se). Factor group analysis yields:

\begin{eqnarray*}
(C_{2v}): \Gamma = A_g + B_{2g} +B_{3g}  + B_{1u} + B_{2u} + B_{3u},
\end{eqnarray*}
\begin{eqnarray*}
(C_s): \Gamma = 2A_g + B_{1g} + 2B_{2g} +B_{3g} + A_u+ 2B_{1u} + B_{2u} + 2B_{3u}.
\end{eqnarray*}

Summarizing these representations and subtracting the acoustic (B$_{1u}$+B$_{2u}$+B$_{3u}$) and silent (2A$_u$) modes, we obtained the following irreducible representations of BaFe$_2$Se$_2$O vibrational modes:

\begin{eqnarray*}
\Gamma _{BaFe_2Se_2O}^{optical}
=6A_{g}+2B_{1g}+6B_{2g}+4B_{3g}+5B_{1u}+3B_{2u}+5B_{3u}
\end{eqnarray*}

Thus 18 Raman and 13 infrared active modes are expected to be observed in the BaFe$_2$Se$_2$O vibrational spectra. Because the crystals have the (001)-orientation (the crystallographic \emph{c}-axis is perpendicular to the plane of the single crystal), we were able to obtain only the A$_{g}$ and B$_{1g}$ symmetry modes in the Raman experiment. The results of the lattice dynamics calculations, together with the experimental data are presented in Table I. Normal modes of the A$_{g}$ symmetry Raman active phonons are presented in Figure 2.

The polarized Raman spectra of BaFe$_2$Se$_2$O, measured from (001)-plane at room temperature and 15 K, for the parallel and crossed polarization configurations, are given in Figure 3 and Figure 4, respectively. The spectra measured for parallel polarization configurations consist of the A$_g$ symmetry modes. Four modes at about 64, 96, 163, and 227 cm$^{-1}$ (300 K) are clearly observed for the $c(bb)\bar{c}$ polarization configuration and two additional modes at about 250 cm$^{-1}$ for the $c(aa)\bar{c}$ polarization configuration. For the crossed $c(ab)\bar{c}$ polarization configuration (see Fig. 4) two Raman active B$_{1g}$ symmetry modes at 91 and 177 cm$^{-1}$ are observed. Vertical bars in Figs. 3 and 4 denote the calculated energies of the A$_g$ and B$_{1g}$ symmetry mode, which are in rather good agreement with experimentally observed ones.

\begin{table}[ht]
\caption{Calculated and experimentally observed values of Raman active phonon mode energies (in cm$^{-1}$) of BaFe$_2$Se$_2$O single crystal.} % title of Table
\renewcommand{\arraystretch}{0.6}
\centering  % used for centering table
\begin{tabular}{c c r r c c c c c} % centered columns (8 columns)
\hline\hline
\multicolumn{5}{c}{Raman modes}&\vline&\multicolumn{3}{c}{Infrared modes} \\  \hline                   %inserts double horizontal lines
Symmetry & Calc. & \multicolumn{2}{c}{Experiment}& Activity & \vline&Symmetry&Calc.&Activity\\ [0.5ex] % inserts table
&&300 K &15 K&&\vline&&&\\%heading

\hline

A$^1_{g}$ & 58.2 & 64 &68 &(xx, yy, zz)&\vline&A$^1_{u}$ & 66.8 & silent\\
A$^2_{g}$ & 110.0 & 96 &99.6 &"&\vline&A$^2_{u}$ & 176.7 &"\\
A$^3_{g}$ & 162.5 & 163 &171.7 &"&\vline&-----&-----&-----\\
A$^4_{g}$ & 242.0 & 227 &232.6 &"&\vline& B$^1_{1u}$ & 73.4 & \\
A$^5_{g}$ & 249.2 & 245 &253 &"&\vline&B$^2_{1u}$ & 134.2 &\\
A$^6_{g}$ & 342.0 &  261 &266.7&"&\vline& B$^3_{1u}$ & 229.6 &E$||$c\\

B$^1_{1g}$ & 96.7 & 91 &93 &(xy)&\vline&B$^4_{1u}$ & 260.8 &\\
B$^2_{1g}$ & 170.4 & 177& 181 &"&\vline&B$^5_{1u}$ & 348.6 &\\

B$^1_{2g}$ & 63.8 && &(xz)&\vline&-----&-----&-----\\
B$^2_{2g}$ & 106.6& &&"&\vline& B$^1_{2u}$ & 75.8 & \\
B$^3_{2g}$ & 136.7 &&&"&\vline&B$^2_{2u}$ & 153.2 &E$||$b \\
B$^4_{2g}$ & 180.4 &&&"&\vline&B$^3_{2u}$ & 232.2 &\\
B$^5_{2g}$ & 255.2 & &&"&\vline&-----&-----&-----\\
B$^6_{2g}$ & 663.2& &&"&\vline& B$^1_{3u}$ & 50.0 & \\

B$^1_{3g}$ & 58.7 & &&(yz)&\vline&B$^2_{3u}$ & 113.5\\
B$^2_{3g}$ & 88.9 &&&"&\vline& B$^3_{3u}$ & 183.0 & E$||$a\\
B$^3_{3g}$ & 171.5 && &"&\vline&B$^4_{3u}$ & 281.2 &\\
B$^4_{3g}$ & 264.1 &&&"&\vline& B$^5_{3u}$ & 662.5 & \\

\hline\hline
\end{tabular}
\label{table:nonlin} % is used to refer this table in the text
\end{table}

\begin{figure}
\includegraphics[width=0.6\textwidth]{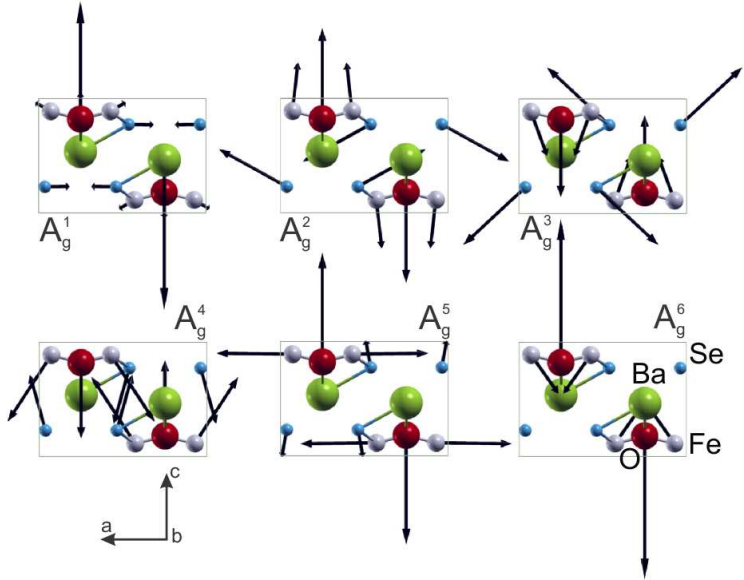}
\caption{(Color online) Normal modes of the A$_g$ symmetry Raman active vibrations of BaFe$_2$Se$_2$O projected to the (ac) plane. The arrow lengths are proportional to the vibrational forces.}
\label{fig2}
\end{figure}

\begin{figure}
\includegraphics[width=0.6\textwidth]{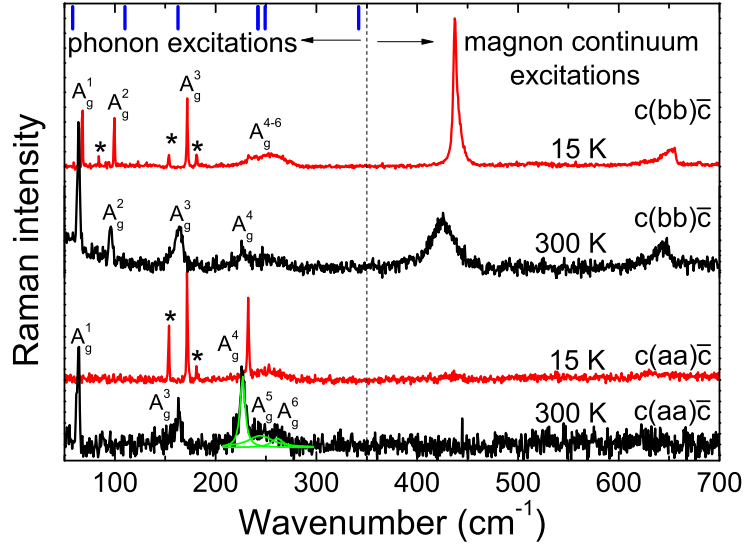}
\caption{(Color online) The (aa) and (bb) polarized Raman scattering spectra of BaFe$_2$Se$_2$O single crystals measured at room temperature and at 15 K. Vertical bars are calculated values of the A$_g$ symmetry Raman active vibrations. Asterisks denote new Raman modes which appear at temperatures below $T_N$=240 K.  }
\label{fig3}
\end{figure}

\begin{figure}
\includegraphics[width=0.7\textwidth]{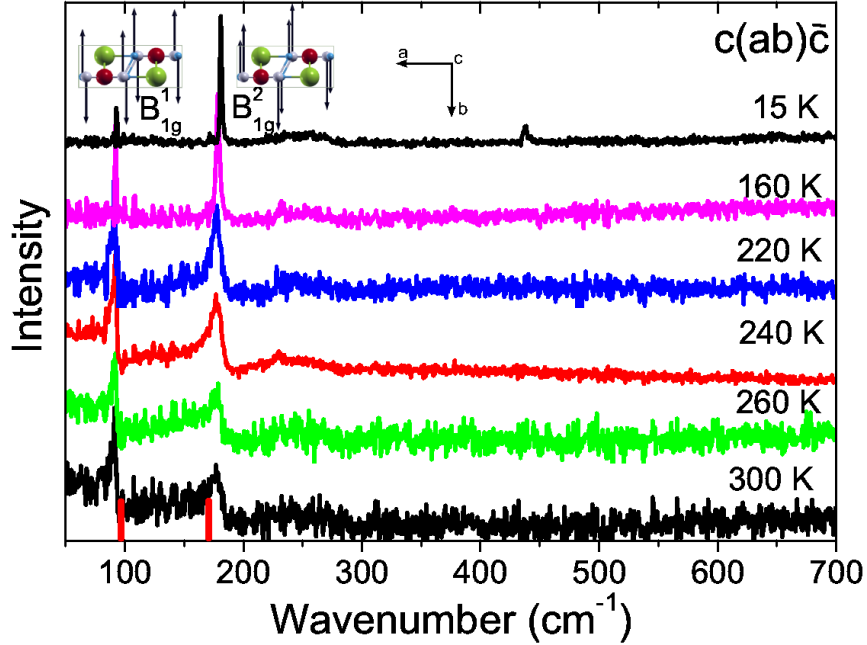}
\caption{(Color online) The (ab) polarized Raman scattering spectra of BaFe$_2$Se$_2$O single crystals measured at various temperatures between 15 K and 300 K. Vertical bars are calculated values of the B$_{1g}$ symmetry modes. Insets show the normal modes of the B$_{1g}^1$ and B$_{1g}^2$ symmetry Raman active vibrations, which originate from the Se and Fe atoms vibrations along the b-axis, respectively.}
\label{fig4}
\end{figure}

\begin{figure}
\includegraphics[width=0.6\textwidth]{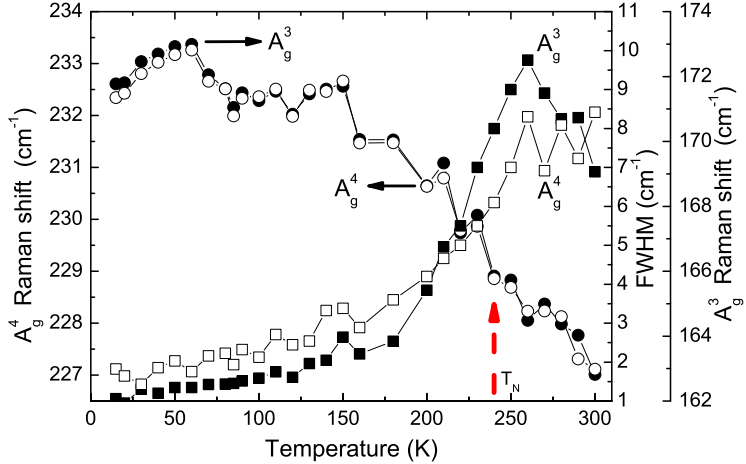}
\caption{(Color online)  Energy (circles) and linewidth (squares) temperature dependence of the A$_g^3$ and A$_g^4$ symmetry modes.}
\label{fig5}
\end{figure}

\begin{figure}
\includegraphics[width=0.6\textwidth]{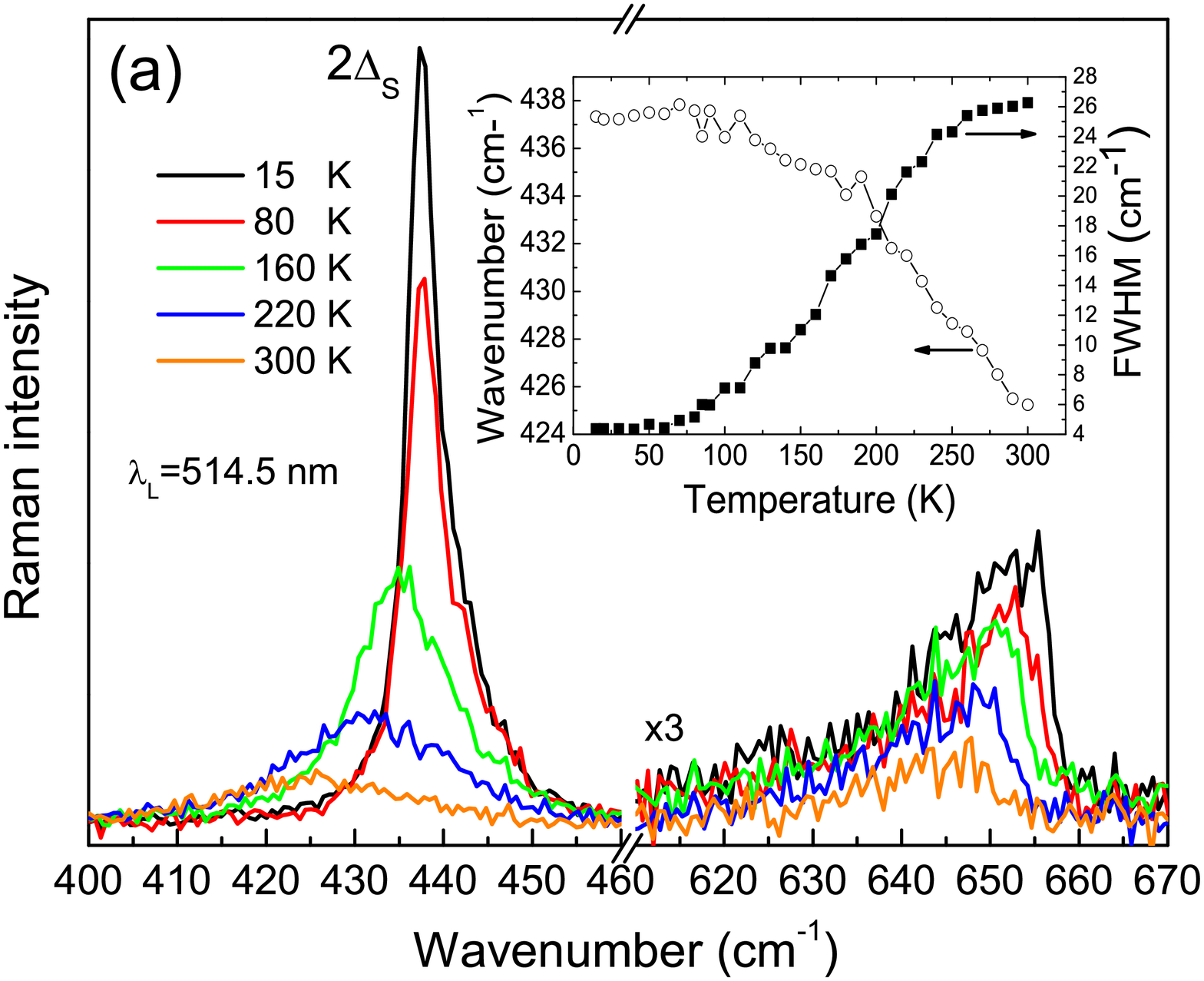}
\includegraphics[width=0.6\textwidth]{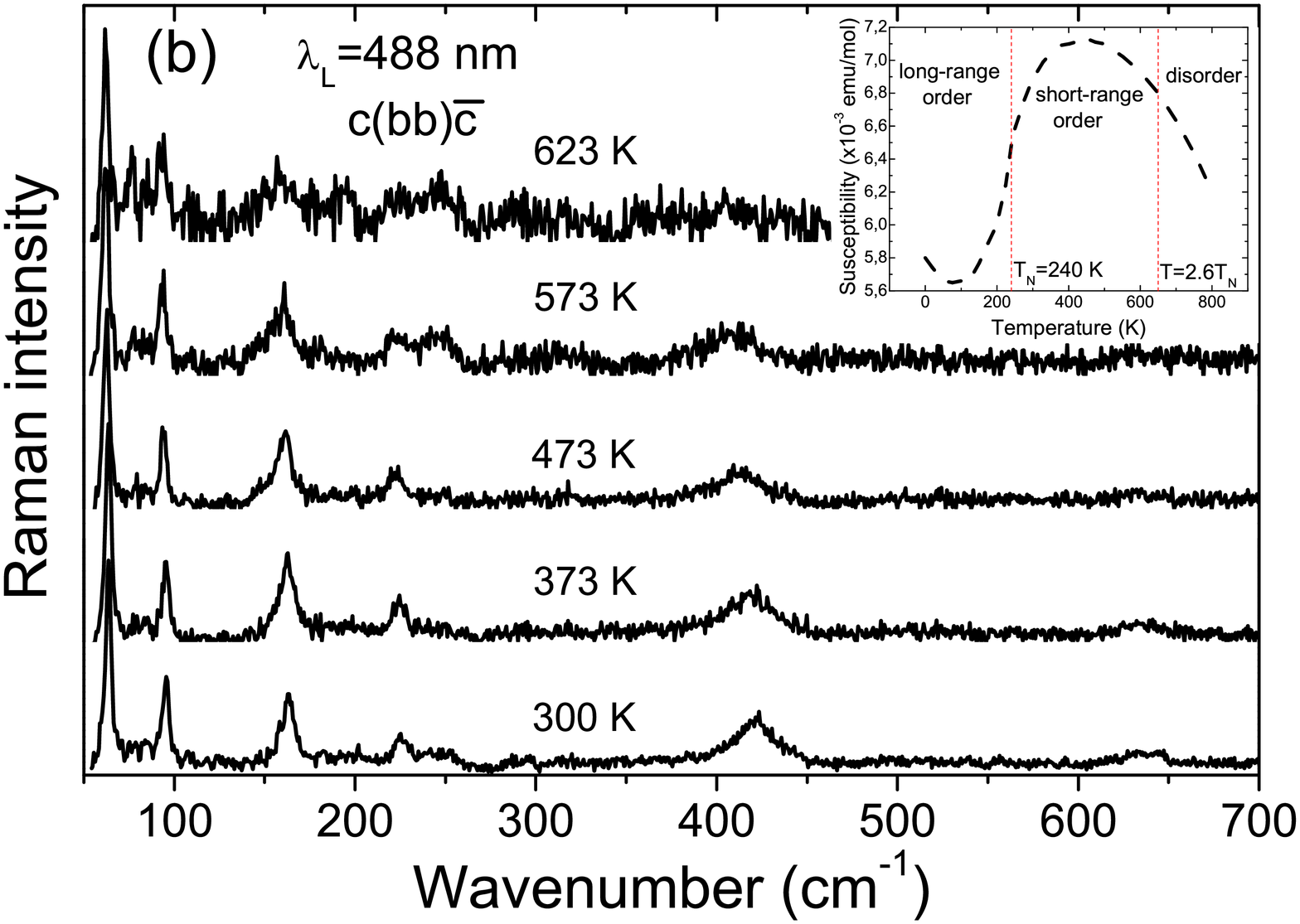}
\caption{(Color online) a) The (bb) polarized Raman scattering spectra of BaFe$_2$Se$_2$O single crystals measured at various temperatures between 15 K and 300 K in the 400-670 cm$^{-1}$ spectral range. Inset: Energy (circles) and linewidth (squares) of the 2$\Delta_S$ (spin gap) mode as a function of temperature. b) The same Raman spectra measured in the temperature range between 300 K and 623 K. Inset: Magnetic susceptibility vs temperature from Ref. \onlinecite{Han}.}
\label{fig6}
\end{figure}

As we have shown in Fig. 2, the lowest energy A$_g^1$ mode is dominated by Ba atom vibrations along the \emph{c}-axis, the A$_g^2$ mode represents vibrations of Fe, Se and O atoms which tend to rotate FeSe$_3$O chains around of the \emph{b}-axis. The A$_g^3$ mode represents dominantly Se atom vibrations, whereas the A$_g^4$ mode originates dominantly from Fe atom vibrations. The A$_g^5$ mode represents vibration of Fe atoms along the \emph{a}-axis together with vibrations of the oxygen atoms along the \emph{c}-axis. Finally, the A$_g^6$ mode originates from oxygen vibration along the \emph{c}-axis. Intensities of the last two modes are rather small as can be seen in Fig. 3, most probably because of oxygen deficiency in the sample. In fact, according to M\"{o}ssbauer spectroscopy measurements\cite{Lei} about 8\% of our sample shows FeSe$_4$ tetrahedral coordination of Fe ions instead of FeSe$_3$O. This suggest the possible presence of vacancies on the O site, excess Se and  BaFe$_2$Se$_{2+\delta}$O$_{1-\delta}$ stoichiometry, which can alter intensity of oxygen related vibrational modes. Energies of the A$_g^5$ and A$_g^6$ modes are obtained from a broad structure peaked at about 250 cm$^{-1}$ using lorentzian lineshape deconvolution technique (see room temperature spectra for the (aa) polarization, Fig. 3).

By lowering the temperature A$_g$ modes increase their energies and sharply reduce the linewidth below the phase transition temperature $T_N$=240 K, as it is shown for the A$_g^3$ and A$_g^4$ modes in Figure 5. Besides that signal to noise ratio in the Raman spectra measured at 15 K is significantly reduced by increasing temperature to 300 K. We believe that this change is related to the spin ordering. In addition, new modes appear at temperatures below 240 K, most probably as a consequence of crystal structure and/or crystal symmetry change at $T_N$. These modes are denoted with asterisk in Fig. 3.

In addition to the phonon modes, two asymmetric structures are observed at about 425 and 645 cm$^{-1}$ at the room temperature Raman scattering spectrum for the $c(bb)\bar{c}$ polarization configuration only. These structures increase their intensity and harden substantially by lowering the temperature, remaining asymmetric up to the lowest temperature of 15 K, see Figure 6 (a). Because the optical phonon region for the A$_g$ symmetry modes is below 350 cm$^{-1}$ we assigned these modes as magnetic ordering related from the following reasons:

\emph{(i)} The magnetic susceptibility measurements\cite{Han} show a broad peak between 400 K and 500 K with a maximum around $T_{max}$=450 K and an abrupt decrease towards zero at $T_N$=240 K. This shape of the magnetic susceptibility is observed in many two-dimensional magnetic systems \cite{Lemmens} including the parent compounds of iron-based superconductors\cite{Rotter} and is connected to the short-range correlation of the local moments\cite{Zhang}. The ratio of $T_N$/$T_{max}$ is usually used to estimate the extent of the low-dimensional magnetic correlations\cite{He}. As $T_N$/$T_{max}$ in BaFe$_2$Se$_2$O is 0.53, it can be concluded that BaFe$_2$Se$_2$O is a quasi-two-dimensional magnetic system, or more precisely spin-ladder (dimer) system.

\emph{(ii)} Magnon related continuum appears only in the (bb) polarisation, \emph{i.e.} in the direction where the light polarization is colinear with spin orientation polarization.

\emph{(iii)}  In spin-ladder systems a triplet gap mode and two magnon continuum may appear in the Raman spectra \cite{Sushkov}. As we already mentioned, the magnetic structure of BaFe$_2$Se$_2$O may be considered as the AFM spin-ladder (dimer) structure with non-negligible inter-ladder (dimer) interaction (strong Fe-O-Fe super-exchange interaction $J_1$ along the rungs of the S=2 spin-ladders is dominant; the inter-ladder interaction $J_2$ is more than three times weaker than along the rungs\cite{Han}). Raman scattering spectra of the AFM dimer magnetic structure was a subject of our interest in Ref. \onlinecite{Milan}, where we have shown that spin dimmer structure in CaV$_2$O$_5$ is represented in the form of two-magnon continuum with the singularities of the one-dimensional density of (two-magnon) states. In the case of BaFe$_2$Se$_2$O the two magnon continuum is well separated from the A$_g$ optical phonon range. The onset of the magnetic continuum appears at about 425 cm$^{-1}$ (437 cm$^{-1}$) at room temperature (15 K) as an asymmetric shape line with a tail towards higher energies (M$_0$ critical point in the 1D density of magnon states). This continuum ends with the $\lambda$-shape line (M$_1$ critical point in the 1D density of magnon states) peaked at about 645 cm$^{-1}$ (655 cm$^{-1}$) at room temperature (15 K). The total width of the continuum is estimated to be around 230 cm$^{-1}$ (15 K). The energy of the 437 cm$^{-1}$ line saturates at temperatures below 100 K, see the inset of Fig. 6.

\emph{(iv)} The well defined peak around 2$\Delta_S$ is expected in the Raman spectra of the spin-ladder (dimer), due to the singularities of the one-dimensional density of two-magnon states\cite{Milan}. Thus, we assigned the onset of the magnon continuum to 2$\Delta_S$= 437 cm$^{-1}$ (15 K) and obtain the spin-gap in BaFe$_2$Se$_2$O to be 219 cm$^{-1}$ (27 meV).

\emph{(v)} There is a common belief that the short-range magnetic order in oxychalcogenides extends at least up to temperature of 2$T_N$.\cite{He} In order to check this assumption we have measured Raman spectra for the (bb) polarization at high temperatures up to 350 $^0$C (623 K). As it can be seen from Fig 6 (b) the two magnon continuum singularity modes decrease their intensity in comparison to phonon modes and at temperature of about 623 K these modes completely disappear. Accordingly, we concluded that the short-range magnetic order in BaFe$_2$Se$_2$O persist up to about 2.6$T_N$.

Finally, we believe that in the short-range order region [between $T_N$ and 2.6$T_N$, see the inset in Fig. 6 (b)] there are also antiferromagnetic fluctuations of local magnetic moments. Spin fluctuations are coupled with the B$_{1g}$ symmetry phonon modes. This coupling is manifested in the appearance of the B$_{1g}$ mode line shape  asymmetry (the Fano-profile shape). Below $T_N$ (see Fig. 4), this asymmetry vanishes as a system enters the long-range ordered antiferromagnetic state.

\section{Conclusion}

We have measured the polarized Raman scattering spectra of the BaFe$_2$Se$_2$O single crystals at various temperatures. All Raman-active modes from (ab) plane predicted by factor-group analysis have been experimentally observed and assigned. Frequencies of these modes are in rather good agreement with the lattice dynamics calculations. The A$_{g}$ and B$_{1g}$ mode linewidths and energies change substantially at temperatures below $T_N$=240 K, where system becomes antiferromagneticaly long-range ordered. At frequencies higher than the A$_{1g}$ optical phonon range we have observed magnon continuum, which is related to the AFM dimer structure. Spin gap value of 27 meV is obtained from the position of the onset of the magnetic continuum. High temperature Raman scattering measurements shows that the short-range magnetic order disappears at about 2.6$T_N$. In the short-range order region we have observed coupling of antiferromagnetic spin fluctuations with the B$_{1g}$ symmetry phonons. It would be interesting to test and compare our results of the magnetic correlations with low temperature neutron scattering study of the BaFe$_2$Se$_2$O crystal, as well as, theoretical calculations of magnon dispersions in the S=2 spin ladder (dimer) barium iron oxyselenide compound.

\section{Acknowledgments}
This work was supported by the Serbian Ministry of Education, Science and Technological Development under Projects ON171032 , ON171017 and III45018. Work at Brookhaven was supported by the US DOE under Contract No. DE-AC02-98CH10886 and in part by the Center for Emergent Superconductivity, an Energy Frontier Research Center funded by the US DOE, Office for Basic Energy Science (H.L. and C.P.). Numerical simulations were run on the AEGIS e-Infrastructure, supported in part by FP7 projects EGI-InSPIRE, PRACE-1IP and HP-SEE.

$^{\S}$Present address: Frontier Research Center, Tokyo Institute of Technology, 4259 Nagatsuta, Midori, Yokohama 226-8503, Japan.

%\bibliography{Reference}
%merlin.mbs apsrev4-1.bst 2010-07-25 4.21a (PWD, AO, DPC) hacked
%Control: key (0)
%Control: author (8) initials jnrlst
%Control: editor formatted (1) identically to author
%Control: production of article title (-1) disabled
%Control: page (0) single
%Control: year (1) truncated
%Control: production of eprint (0) enabled
%

\end{document}